\documentclass[11pt]{article}
\usepackage{fullpage}
\usepackage{latexsym}
\usepackage{amsfonts}
\usepackage{url}
\usepackage{theorem}
\usepackage{url}
\usepackage[breaklinks,bookmarks]{hyperref}

\newcommand{\bra}[1]{\langle #1|}
\newcommand{\ket}[1]{|#1\rangle}
\newcommand{\braket}[2]{\langle #1|#2\rangle}
\newcommand{\norm}[1]{\| {#1} \|}
\newcommand{\OO}[1]{\mathrm{O}\mathchoice{\!}{\!}{}{}\left({#1}\right)}
\renewcommand{\O}[1]{\mathrm{O}({#1})}
\newcommand{\Om}[1]{\Omega(#1)}
\newcommand{\Th}[1]{\Theta\mathchoice{\!}{\!}{}{}\left(#1\right)}
\newcommand{\GF}[1]{\mathbb{GF}(#1)}
\newcommand{\poly}{\mathrm{poly}}
\long\def\rem#1{}
\def\01{\{ 0, 1 \}}

\renewcommand{\Pr}[2][]{\mathrm{Pr}#1[#2]}
\newcommand{\Exp}[2][]{\mathrm{E}#1[#2]}
\newcommand{\given}{\mathop{|}}
\newcommand{\eps}{\varepsilon}
\newcommand{\comb}[2]{( { #1 \atop #2 } )}
\newcommand{\nonempty}{{W \cap R \times S \ne \emptyset}}
\newcommand{\Va}{\mathbf{a}}
\newcommand{\Vb}{\mathbf{b}}
\newcommand{\Vp}{\mathbf{p}}
\newcommand{\Vq}{\mathbf{q}}
\newcommand{\PV}{\mathrm{PV}}
\newcommand{\VO}{{\em Verify Once}}
\newcommand{\VF}{{\em Verify Full}}

\newcommand{\epsrs}{{\frac {r s} {n^2}}}

\newtheorem{Theorem}{Theorem}
\newtheorem{Lemma}[Theorem]{Lemma}
\newenvironment{Proof}[1][Proof.]{
        \par
        \noindent \textbf{#1}
}{
        \nobreak\leavevmode
        \hfill $\Box$\par\medskip
}
\newenvironment{Remark}[1][Remark.]{
        \par\medskip
        \noindent {\em #1}
}{
        \par\medskip
}
\newenvironment{Problem}[1][Problem.]{
        \par\medskip
        \noindent {\em #1}
}{
        \par\medskip
}
\newenvironment{shortenum}
{\begin{enumerate}\setlength{\itemsep}{0in}}
{\end{enumerate}}
\newenvironment{shortitem}
{\begin{itemize}\setlength{\itemsep}{0in}}
{\end{itemize}}

\begin{document}

\title{Quantum Verification of Matrix Products}
\author{Harry Buhrman\thanks{%
CWI and University of Amsterdam;
supported in part by the EU fifth framework project QAIP, IST-1999-11234,
and RESQ, IST-2001-37559, and an NWO grant.}\\
buhrman@cwi.nl
\and
\addtocounter{footnote}{-1}%
Robert \v Spalek\footnotemark\\
sr@cwi.nl}
\date{
}
\maketitle

\begin{abstract}
We present a quantum algorithm that verifies a product of two $n
\times n$ matrices over any integral domain with bounded error in worst-case
time $\O{n^{5/3}}$ and expected time $\O{n^{5/3} / \min(w, \sqrt
n)^{1/3}}$, where $w$ is the number of wrong entries.  This improves
the previous best
algorithm~\cite{AmbainisBuhrmanHoyerKarpinskyKurur02} that runs in
time $\O{n^{7/4}}$.  We also present a quantum matrix multiplication
algorithm that is efficient when the result has few nonzero entries.
\end{abstract}

\section{Introduction}
The computational complexity of matrix multiplication is a subject of
extensive study. Matrix multiplication is the central algorithmic
part of many applications like for example solving linear systems of
equations and computing the transitive closure. A fast algorithm for
matrix multiplication thus implies a fast algorithm for a variety of
computational tasks. Strassen~\cite{strassen69} was the first to show
that surprisingly two $n \times n$ matrices can be multiplied in time
$n^{2+\alpha}$ for $\alpha < 1$. His result was improved by many
subsequent papers. The best known bound to date is an algorithm with
$\alpha \approx 0.376$ by Coppersmith and
Winnograd~\cite{cw:matrix-mult}. It is a main open problem to
determine the true value of $\alpha$. Freivalds
showed~\cite{Freivalds79} that {\em verifying} whether the product of two $n
\times n$ matrices is equal to a third can be done with high
probability in time proportional to $n^2$. We will refer to this
latter problem as matrix verification.

We study the computational complexity of matrix multiplication and
verification on a quantum computer. The first to study matrix verification in
the quantum mechanical setting were Ambainis, Buhrman, H\o yer, Karpinski, and
Kurur~\cite{AmbainisBuhrmanHoyerKarpinskyKurur02} who used a clever
recursive version of Grover's algorithm to verify whether two $n\times n$
matrices equal a third in time $\O{n^{7/4}}$, thereby improving the
optimal classical bound of Freivalds.

In this paper we will construct a bounded error quantum algorithm for
the matrix verification problem that runs in time $\O{n^{5/3}}$. Suppose we
are verifying whether $A\times B =C$. When the
number of ``wrong'' entries in $C$ is $w$, our
algorithm runs in expected time $\O{n^{5/3} / \min(w, \sqrt
n)^{1/3}}$. For $w=\sqrt{n}$ we have a matching lower bound.

Our algorithm uses the quantum random walk formalism by
Szegedy~\cite{szegedy:q-walk} that he developed as a generalization
of the quantum random walk technique of
Ambainis~\cite{ambainis:eldist}. Ambainis used a quantum random walk
to obtain an optimal quantum algorithm for the element distinctness
problem. If one were to adapt that method directly to the setting of
matrix verification, one does obtain a $\O{n^{5/3}}$ algorithm in terms
of queries to the input. However that algorithm still requires $\Om{n^2}$
time, because it computes several times a matrix product of sub-matrices
that are loaded into the memory.  This costs no additional queries, but it will take
additional time.  The rest of the paper is devoted to improve the time
complexity of the quantum algorithm to $\O{n^{5/3}}$.

We perform a quantum
random walk on the product of two Johnson graphs, analyze its
spectral gap, that is the second smallest eigenvalue of its Laplacian, and estimate that
in our setting enough of the nodes are marked if $A\times B \ne C$.
See Section~\ref{sec:alg} for a detailed description of our
algorithm. We next introduce a combinatorial tool to analyze the
behavior of our algorithm when many of the entries are wrong.
Finally we use our fast quantum matrix verification algorithm as a
building block to construct a quantum algorithm for computing the
actual matrix product $A \times B$ that is substantially faster than
any known classical method, when there are not too many non-zero
entries in the final product.

\section{Preliminaries}

\subsection{Quantum query complexity}

We assume familiarity with quantum computing~\cite{nielsen&chuang:qc} and
sketch the model of quantum query complexity.  Suppose we want to compute some
function $f$.  For input $x\in\01^N$, a \emph{query} gives us access to the
input bits. It corresponds to the unitary transformation
\[
O:\ket{i,b,z}\mapsto\ket{i,b\oplus x_i,z}.
\]
Here $i\in\{1,\ldots,N\}$ and $b\in\01$; the $z$-part corresponds to the
workspace, which is not affected by the query.  We assume the input can be
accessed only via such queries.  A $t$-query quantum algorithm has the form
$A=U_t O U_{t-1}\cdots O U_1 O U_0$, where the $U_k$ are fixed unitary
transformations, independent of $x$.  This $A$ depends on $x$ via the $t$
applications of $O$.  The algorithm starts in the initial state
$\ket{0^k}$ and its \emph{output} is the result of measuring a dedicated part of
the final state $A\ket{0^k}$, where $k$ is the total amount of space used by
the algorithm.

\subsection{Quantum search}

One of the most interesting quantum algorithms is Grover's search
algorithm~\cite{grover:search,bbht:bounds}.  It can find an index of an input
bit $x_i$ in an $n$-bit input such that $x_i = 1$ in expected number of
$\O{\sqrt{n/(|x|+1)}}$ queries,
where $|x|$ is the Hamming weight (number of ones) in the input.
Grover's algorithm can be cast in more general terms as amplitude
amplification: given a quantum algorithm $A$ that accepts with probability
$p$, then it can be amplified to have constant success probability with
$\sqrt{1/p}$ iterations of $A$.

Given $n$ numbers $x_1, \dots, x_n$ as input, the element distinctness problem
is the task to determine whether there are two distinct indices $i$ and $j$
such that $x_i = x_j$.  Ambainis in a very nice paper \cite{ambainis:eldist}
applied quantum random walks in a novel way and constructed a quantum
algorithm that solves element distinctness in $\O{n^{2/3}}$ queries.  This
algorithm is faster than the
algorithm~\cite{betal:distinctness} which is based on amplitude amplification
and uses $\O{n^{3/4}}$
queries.  Ambainis's method was generalized by
Szegedy~\cite{szegedy:q-walk} for all graphs and even for all symmetric Markov chains
(with non-uniform transition probabilities).
Szegedy's method can be regarded as a quantum walk version of amplitude
amplification~\cite{bhmt:countingj}.


\subsection{Previous best algorithm for verification of matrix products}

Ambainis et al.~\cite{AmbainisBuhrmanHoyerKarpinskyKurur02} discovered a
quantum algorithm running in time $\O{n^{7/4}}$.  Since it was never
published, we will briefly sketch it here.  Let $A, B, C$ be $n \times n$
matrices.  First, partition the matrices $B$ and $C$ into $\sqrt n$ blocks of
$\sqrt n$ columns each.  It
holds that $A B = C$ iff $A B_i = C_i$ for every $i$, where $B_i$ and $C_i$
are the sub-matrices of size $n \times \sqrt n$.  The verification of $A
B_i = C_i$ can be done with bounded error in time $\O{n^{3/2}}$ as follows:
choose a random vector $x$ of length $\sqrt n$, multiply both sides of the
equation by $x$ from the right side, compute classically $y = B_i x$ and $z =
C_i x$, and verify the matrix-vector product $A y = z$ by a Grover search.
The search over $n$ rows takes $\O{\sqrt n}$ iterations and a verification of
one row takes time $n$.  Now, we apply amplitude
amplification on the top of this sub-routine $V_i$, and compute the And of all $\sqrt n$
blocks using $n^{1/4}$ calls to $V_i$.

\subsection{Notation}

Let $[n]$ denote the set $\{ 1, 2, \dots, n \}$.
Let $A_{n \times m}$ denote a matrix $A$ of dimension $n \times m$.  Let
$A^T$ denote the transpose of $A$.  For a $R \subseteq [n]$, let $A|_R$ denote
the $|R|\times m$ sub-matrix of $A$ restricted to the rows from $R$.
Analogously, for every $S \subseteq [m]$, let $A|^S$ denote the $n \times |S|$
sub-matrix of $A$ restricted to the columns from $S$.
Let $\lambda(M)$ denote the \emph{spectral norm} of a matrix $M$; it is equal
to the largest eigenvalue of $M$ for symmetric $M$.
For a set $S$, let $\comb S k$ denote all subsets of $S$ of size $k$.
An \emph{integral domain} is a commutative ring with identity and no divisors
of 0.  

For a graph $G$, let $V_G$ denote the vertices of $G$ and let $E_G$ denote the
edges of $G$.  The normalized \emph{Laplacian matrix} ${\cal L}(G)$ of an
undirected graph $G$ is a symmetric $|V_G| \times |V_G|$ matrix defined by
${\cal L}_{i,j}(G) = 1$ if $i = j$, it is $-{1 / \sqrt{d_i d_j}}$ if $i \ne
j$, and $0$ otherwise.  A \emph{spectral gap} of a graph $G$, often called the
\emph{Fiedler value} of $G$, equals to the second smallest eigenvalue of
${\cal L}(G)$; it is nonzero if $G$ is connected.  The \emph{Johnson graph
$J(n,k)$} is defined as follows: its vertices are subsets of $[n]$ of size
$k$, and two vertices are connected iff they differ in exactly one number.
Let $G = G_1 \times G_2$ denote the \emph{graph categorical product} of two
graphs $G_1, G_2$, defined as follows: $V_G = V_{G_1} \times V_{G_2}$, and
$((g_1, g_2), (g_1', g_2')) \in E_G$ iff $(g_1, g_1') \in E_{G_1}$ and $(g_2,
g_2') \in E_{G_2}$.

\section{Algorithm for verification of matrix products}
\label{sec:alg}

Let $A, B, C$ be $n \times n$
matrices over any integral domain.  A \emph{verification of a matrix product} is
deciding whether $A B = C$.  We construct an efficient quantum walk algorithm for
this problem.  It is described in
Figure~\ref{fig:verify} and
its expected running time is analyzed in Sections~\ref{sec:analysis-verify}
and \ref{sec:lowerb-zeta}.

\begin{figure}[htbp]
\begin{center}
\framebox{
\begin{minipage}{0.9\hsize}
\noindent
{\em Product Verification}
(input size $n$, matrices $A, B, C$)
returns 1 when $A B \ne C$:
\begin{enumerate}
\item
Take any $1 < \lambda < \frac87$, for example $\lambda = \frac{15}{14}$.
\item
For $i = 0, 1, \dots, \log_\lambda (n^{2/3}) + 9$, repeat 16 times the following:
\begin{itemize}
\item
Run \VO{} $(\sqrt[3] 8 \cdot \lambda^i)$.
\item
If it returns 1, then return ``not equal''.
\end{itemize}

\item
Return ``equal''.
\end{enumerate}
\end{minipage}
}
\vskip 10pt
\framebox{
\begin{minipage}{0.9\hsize}
\noindent
\VO{}
(number of rows $k$)
returns 1 when $A B \ne C$ is detected:
\begin{enumerate}
\setcounter{enumi}3
\item
Pick the number of iterations $\ell$ uniformly at random from
$\{ 1, 2, \dots, k \}$. \\
Pick a random row vector $\Vp$ and a random column vector $\Vq$ of length $n$.
\item
{\bf Initialization.}
Put the quantum register into superposition
\(
\sum_{R \subseteq [n]\atop |R|=k} \sum_{S \subseteq [n]\atop |S|=k} \ket{R} \ket{S}.
\)
(Think of $R$ as a subset of the rows of $A$ and $S$ as a subset of the
columns of $B$.)
Compute $\Va_R = \Vp|^R \cdot A|_R$, $\Vb_S = B|^S \cdot \Vq|_S$, and $c_{R,S}
= \Vp|^R \cdot C|_R^S \cdot \Vq|_S$ in time $2 k n + k^2$.  Let $\ket z = \ket
+ = {\ket0 + \ket1 \over \sqrt2}$.  The quantum state is now
\[
\ket + \sum_{R,S} \ket{R, \Va_R}\, \ket{S, \Vb_S}\, \ket{c_{R,S}}.
\]

\item
{\bf Quantum walk.}
Conditioned on $\ket z$, perform $\ell$ iterations of the following:
\label{it:qwalk}
\begin{enumerate}
\item
{\bf Phase flip.}
Multiply the quantum phase by $-1$ iff $\Va_R \cdot \Vb_S \ne c_{R,S}$.
The scalar product is verified in time $n$ using no queries.

\item
{\bf Diffusion.}
Perform one step of quantum walk on $(R,S)$, that is exchange one row and one
column.  The update of $\Va_R$, $\Vb_S$, and $c_{R,S}$ costs $2 n$ queries to
$A$, $2 n$ queries to $B$, and $4 k$ queries to $C$.
\end{enumerate}

\item
Apply the Hadamard transform on $\ket z$, measure it, and return the outcome.
\end{enumerate}
\end{minipage}
}
\end{center}
\caption{Quantum algorithm for verification of matrix products}
\label{fig:verify}
\end{figure}

The basic outline is the following: {\em Verify Once} estimates the scalar
product of the superposition computed by the quantum walk and the uniform
superposition.  The inequality $\Va_R \cdot \Vb_S
\ne c_{R,S}$ can only be true when $A|_R \cdot B|^S \ne C|_R^S$, that is when
$C|_R^S$ contains at least one wrong entry.  If $A B = C$, then the quantum
walk does nothing, because the phase flip is never performed and the diffusion
on a uniform superposition is equal to identity.  Hence the superposition
computed by the quantum walk stays uniform, and the measurement of $\ket z$ always
yields $0$.  On the other hand, if $A B \ne C$ and $k$ is sufficiently large,
then for $\ell$ drawn uniformly from $\{ 1, 2, \dots, k \}$, with high
probability, the quantum walk converges to a superposition almost orthogonal
to the uniform superposition and the measurement of $\ket z$ yields 1 with
probability close to $\frac12$.  The
loop in {\em Product Verification} tries a sequence of exponentially
increasing $k$.  The idea of multiplying the matrices in \VO{} from both sides
by random vectors $\Vp, \Vq$ is explained in Section~\ref{subsec:analysis}.
It allows us to achieve both a better running time and smaller space complexity.

\section{Analysis of the algorithm}
\label{sec:analysis-verify}

In this section, we prove that {\em Product Verification} has one-sided
bounded error and estimate its expected running time depending on the set
of wrong entries.  We use a recent result by Szegedy~\cite{szegedy:q-walk}, which can
be regarded as a quantum walk version of quantum amplitude
amplification~\cite{bhmt:countingj}.
Its proof is outlined in Appendix~\ref{app:qwalk}.

\begin{Theorem}
[Szegedy~\cite{szegedy:q-walk}]
\label{th:q-walk}
Let $G$ be an undirected graph on vertex set $X$,
and let $\delta_G$ be the spectral gap of $G$.  Some set of vertices $M \subseteq
X$ are marked with the promise that $|M|$ is either zero or at least $\eps
|X|$.  For every $m = \Om{1 / \sqrt{\delta_G \eps}}$, the following quantum
algorithm decides whether $M$ is non-empty with one-sided error $\gamma \le \frac78$
in time $\O{T_X + m \cdot (T_M + T_G)}$:
\begin{shortenum}
\item
Initialization.
Compute a uniform superposition over $X$; let the time be $T_X$.
\item
Pick $1 \le \ell \le m$ uniformly at random.  Repeat $\ell$ times the following: \\
(1) Phase flip.
Flip the quantum phase if an element is marked; let the time be $T_M$. \\
(2) Diffusion.
Perform one step of quantum walk on $G$; let the time be $T_G$.
\item
Estimate the scalar product of the quantum walk distribution and the
uniform distribution.
\end{shortenum}
\end{Theorem}

\subsection{Analysis of \emph{Product Verification}}
\label{subsec:analysis}

We analyze the expected running time of the algorithm as follows.  Let \VF{}
denote a modified version of \VO{} that does not use the random vectors $\Vp,
\Vq$, but instead reads all the sub-matrices into memory and verifies $A|_R
\cdot B|^S = C|_R^S$ in the phase-flip step (\ref{it:qwalk}a).  \VF{} has the
same query complexity as \VO{}, but its space complexity and running time are
bigger.  (Although the phase-flip step (\ref{it:qwalk}a) costs no additional
queries, the time needed to compute classically $A|_R \cdot B|^S$ is at least
$k n$, whereas the scalar product $\Va_R \cdot \Vb_S$ can be computed in time
$n$.%
\footnote{It seems that this slowdown ``time $\gg$ \#queries'' is a typical
property of algorithms using quantum walks: the quantum algorithm for element
distinctness~\cite{ambainis:eldist} needs to use random hash functions to
remedy it, and it is open whether triangle finding~\cite{mss:triangle} can be
improved this way.})
We analyze the error of \VF{}, because the multiplication by $\Vp, \Vq$ complicates
the analysis.  For example, if there is exactly one
wrong entry and the multiplication is over $\GF 2$, then with probability
$\frac 3 4$, the wrong entry is completely hidden by multiplication by zero.
However, we prove the following statement:

\begin{Lemma}
\label{lem:random-vector}
Let $A B \ne C$.  The probability that \VO~$(\sqrt[3] 8 \cdot k)$ outputs 1 at
least once in 16 independent trials, each time with new random $\Vp, \Vq$, is
bigger than the success probability of one call to \VF~$(k)$.
\end{Lemma}

Using this lemma, it is sufficient to analyze the error of the algorithm as if
it is performing \VF{} in each step.  Let $W = \{ (i,j) \given (A
B - C)_{i,j} \ne 0 \}$ be the set of wrong entries of the
matrix product and let $R, S \subseteq [n]$ denote subsets of
rows of $A$ and columns of $B$.  We mark $(R,S)$ iff $C|_R^S$ contains a
wrong entry, formally $A|_R \cdot B|^S \ne C|_R^S$, or equivalently
$\nonempty$.
The performance of the algorithm depends on the \emph{fraction of marked
pairs} $\eps(W,k) = \Pr[_{R,S}]{(R,S)\mbox{ is marked}}$, where $|R|=|S|=k$.
In Section~\ref{sec:lowerb-zeta}, we prove the
following lower bound on $\eps(W,k)$:

\begin{Lemma}
\label{lem:eps}
Let $q(W) = \max( |W'|, \min(|W|, \sqrt n))$, where $W'$ is the largest
independent subset of $W$, that is it contains at most one 1 in every row and
column.
For every $W$ and $k \le {n^{2/3} / q(W)^{1/3}}$,
it holds that $\eps(W,k) = \Om{{k^2 \over n^2} q(W)}$.
\end{Lemma}

We also need the following two statements, whose proofs are in
Appendix~\ref{app:qwalk}:

\begin{Lemma}
\label{lem:spectral-gap}
Let $1 \le k \le \frac n 2$.  The spectral gap of $G = J(n, k) \times J(n, k)$
is $\delta_G = \Th{1/k}$.
\end{Lemma}

\begin{Lemma}
\label{lem:test}
Let $\ket \varphi$ and $\ket \psi$ be quantum states,
 let $\ket X = \frac {\sqrt 2} 2 (\ket{0, \varphi} +
\ket{1, \psi})$, and let $\ket Y = (H \otimes I) \ket X$.  If the first
qubit of $\ket Y$ is measured in the computational basis, then $\Pr{Y = 1} =
\frac 1 2 (1 - \braket \varphi \psi)$.
\end{Lemma}


\begin{Theorem}
\label{th:exp-time}
{\em Product Verification} always returns ``equal'' if $A B = C$.  If $A B \ne
C$, then it returns ``not equal'' with probability at least $\frac 2 3$.  Its
worst-case running time is $\O{n^{5/3}}$, its expected running time is
$\O{n^{5/3} / q(W)^{1/3}}$, and its space complexity is $\O n$.
\end{Theorem}

\rem{
We also show that verification of some matrix products with $n$ wrong
entries requires $\Om{n^{3/2}}$ queries, hence {\em Product Verification} is
optimal for $|W| \in [\sqrt n, n]$.
}

\begin{Proof}
By Lemma~\ref{lem:random-vector}, if we replace the calls to \VO{} by \VF{} in
Figure~\ref{fig:verify} and skip repeating each loop 16 times and
multiplication of $k$ by $\sqrt[3] 8$, the success probability is decreased.
Hence if we compute an upper bound on the expected number of iterations of
such an algorithm, it will also hold for the original algorithm.
Let us thus analyze the running time of the original algorithm assuming the error
analysis is of \VF.
\VO{} walks $\ell$ quantum steps on the
graph categorical product of two Johnson graphs $G = J(n,k) \times J(n,k)$.
The marked vertices of $G$ correspond to marked pairs $(R,S)$, that is the
pairs such that $A|_R \cdot B|^S \ne C|_R^S$.  The
initialization costs time $T_X = \O{k n}$, a phase flip costs time $T_M = n$,
and one step of the quantum walk costs time $T_G = 4n + 4k = \O n$.  The running
time of \VO{} is thus $\O{(k + \ell) n} = \O{kn}$.  The scalar product of two
distributions is estimated using Lemma~\ref{lem:test}.

Let $W \ne \emptyset$.  By Theorem~\ref{th:q-walk}, \VO{} recognizes a wrong
matrix product with bounded error for every $m \ge \O{1 / \sqrt{\delta_G\,
\eps(W, k)}}$.  Plug in $\eps(W, k) = \Om{ {k^2 \over n^2} q(W)}$ by
Lemma~\ref{lem:eps} and $\delta_G = \Th{\frac 1 k}$ by
Lemma~\ref{lem:spectral-gap}.  We get that $m \ge \O{n / \sqrt{k q(W)}}$.
In our algorithm, we use $m = k$, which gives the condition
$k \ge k_0 = \O{n^{2/3} / q(W)^{1/3}}$.
Hence for every $k \ge k_0$, \VO{} makes only small one-sided error.  The
algorithm {\em Product Verification} does not know $q(W)$ and $k_0$, but it
instead tries different values of $k$ from the exponentially increasing
sequence $1 \dots \lambda^i$.

The total running time is dominated by the last run.
The expected running time can be written as a telescopic sum $\Exp T = \sum_{t=0}^\infty t
\cdot \Pr{T = t} = \sum_{t=1}^\infty \Pr{T \ge t}$.  {\em Product Verification}
(PV)
calls \VO{} with time $k n = \lambda^i n$ and each call after $k \ge k_0$
fails with probability $\gamma \le \frac78$, hence
\begin{eqnarray*}
\Exp T &=& \sum_i \lambda^i n \cdot \Pr{\mbox{$\PV$ enters the $i$-th loop}}
  \le \sum_{i=0}^{(\log_\lambda k_0) - 1} \lambda^i n
    + \sum_{i=\log_\lambda k_0}^{(\log_\lambda n^{2/3}) + 9}
      \lambda^i n \cdot \gamma^{i - \log_\lambda k_0} \\
  &\le& \O{k_0 n} \left( 1 + \sum_{i=0}^\infty (\lambda \gamma)^i \right)
  = \O{k_0 n}
  = \OO{n^{5/3} \over q(W)^{1/3}},
\end{eqnarray*}
because $\lambda \gamma < \frac 87 \cdot \frac78 = 1$.
The probability that a wrong product is never recognized is
$\le \gamma^9 < \frac13$, where $9$ is the
number of loops after $n^{2/3}$.

$\PV$ never makes an error when $A B = C$.  In this case, the phase flip is
equal to the identity operation.  The diffusion is also equal to the identity on the uniform
distribution, hence the whole quantum walk in \VO{} does nothing and
the qubit $\ket z = \ket +$ is untouched.  Finally, $\PV$ always terminates
when $k \ge \lambda^9 n^{2/3}$, hence its
total running time is $\O{n^{5/3}}$.
\end{Proof}
\goodbreak

It remains to prove Lemma~\ref{lem:random-vector}.
Let us fix random vectors $\Vp, \Vq$.  We call $(R,S)$ \emph{revealing} iff
\(
\Va_R \cdot \Vb_S = (\Vp|^R \cdot A|_R) \cdot (B|^S \cdot \Vq|_S) \ne c_{R,S},
\)
which is equivalent to
\(
\Vp|^R \cdot (A|_R \cdot B|^S) \cdot \Vq|_S \ne \Vp|^R \cdot C|_R^S \cdot \Vq|_S
\)
due to the associativity of matrix multiplication.  As we have already
seen, not every marked pair is revealing.  Let $\zeta_{\Vp,\Vq}(W,k) =
\Pr[_{R,S}]{(R,S) \mbox{ is revealing}}$ denote the \emph{fraction of revealing
pairs}, where $|R| = |S| = k$.  The proof of the following statement is in
Appendix~\ref{app:fraction-revealing}:

\begin{Lemma}
\label{lem:good-vec}
Let $\Vp, \Vq$ be picked uniformly at random.  Then $\Pr{ \zeta_{\Vp,\Vq}(W,k)
\ge \frac 1 8 \eps(W,k) } > \frac 1 8$.
\end{Lemma}

Now, we show that the constant probability of picking good random
vectors is compensated by a constant number of repetitions.
\medskip

\begin{Proof}[Proof of Lemma~\ref{lem:random-vector}]
By Lemma~\ref{lem:good-vec}, the success probability of \VO{} is at least
$\frac p 8$, where $p$ is the success probability of \VO{} given that it
guesses good vectors $\Vp, \Vq$ with $\zeta(W,k) \ge \frac 1 8 \eps(W,k)$.  By
Theorem~\ref{th:q-walk} and the proof of Theorem~\ref{th:exp-time}, $p = 1 -
\gamma$ and $\frac 1 2 \ge p \ge \frac 1 8$ for every $k \ge k_0 = \O{n^{2/3}
/ (q(W) / 8)^{1/3}}$; the factor $\frac 1 8$ in
$\eps(W,k)$ is compensated by taking $\sqrt[3] 8$-times bigger $k$.  The
success probability of 16 independent trials is at least
\(
1 - (1 - \frac p 8)^{16}
  \ge 1 - (e^{-p})^2 \ge 1 - (1 - 0.64 p)^2
  \ge 1.28 p - 0.4 p^2 \ge p,
\)
because $1 - x \le e^{-x}$, $e^{-x} \le 1 - 0.64 x$ for $x \in [0,1]$, and $p
< 0.7$.
\end{Proof}


\subsection{Comparison with other quantum walk search algorithms}

{\em Product Verification} resembles a few other algorithms.  The first
quantum algorithm of this type was the quantum walk algorithm for element
distinctness~\cite{ambainis:eldist}.  The same technique was subsequently
successfully applied to triangle finding~\cite{mss:triangle} and group
commutativity testing~\cite{mn:group-commutativity}.  Both algorithms
walk on the Johnson graph $J(n,k)$.
The analysis of Ambainis~\cite{ambainis:eldist} relies on the
fact that the quantum state stays in a constant-dimensional subspace.  This
constraint is satisfied if there is at most one solution; then the subsets can be
divided into a constant number of cases.  In the non-promise version, the
number of cases is, however, not constant.  The latter papers~\cite{ambainis:eldist,
mss:triangle} solve the non-promise case by projecting the input into a random
subspace.  With high probability, there is exactly one solution in the
subspace; this trick originally comes from Valiant and Vazirani~\cite{vv:uniqsol}.  Since it is
not known whether this technique can be used in more than one dimension, we
solve the non-promise version of product verification using the more general
quantum walk by Szegedy~\cite{szegedy:q-walk} instead of the original one by Ambainis~\cite{ambainis:eldist}.

Theorem~\ref{th:q-walk} is quite general, because it allows walking on an
arbitrary undirected graph.  On the other hand, the algorithm \VO{}
obtained by it is a bit slower than the original Ambainis walk~\cite{ambainis:eldist, mss:triangle}.
First, \VO{} only solves the decision version of the problem and it does not find
the actual position of a wrong entry.  This can be
resolved by a binary search.  Second, \VO{} does the phase flip after every step
of quantum walk instead of doing it once per block of steps.
However, for both element
distinctness and product verification, the additional cost is subsumed by the
cost of the quantum walk.

\section{Lower bounds on the fraction of marked pairs}
\label{sec:lowerb-zeta}

In this section, we try to solve the following combinatorial problem:

\begin{Problem}
Given an $n \times n$ Boolean matrix $W$ and two integers $1 \le r, s \le n$,
what is the probability $\eps(W,r,s)$ that a random $r \times s$ sub-matrix of
$W$ contains a 1?
Or, equivalently:
Given a bipartite graph on $n, n$ vertices, what is the probability that a
randomly chosen induced subgraph with $r, s$ vertices contains at least one
edge?
\end{Problem}

It is simple to prove that $\eps(W,r,s)$ is monotone in all its three parameters.
As we have seen in Theorem~\ref{th:exp-time}, the expected running time of
{\em Product Verification} depends on the fraction of marked pairs, which
is $\eps(W,k,k)$, also denoted there by $\eps(W,k)$.%
\footnote{Our algorithm only tries balanced choices $r = s = k$.
Since the initialization costs $\O{(r+s) n}$, setting one of the variables
smaller does not decrease the query complexity, but it decreases the success
probability of {\em Verify Once}.}
Let us compute $\eps$ when $W$ contains exactly one 1:
\(
\eps(W,r,s)
  = {\comb {n-1} {r-1} \comb {n-1} {s-1}
    / \comb n r \comb n s}
  = {r s \over n^2}.
\)
With this bound, monotonicity, and Theorem~\ref{th:exp-time}, we conclude that {\em
Product Verification} finds the correct answer with bounded error in time
$\O{n^{5/3}}$.  The rest of this section contains a more detailed analysis of
the expected running time of the algorithm for larger $W$.  Unfortunately, we are only
able to prove weak lower bounds on $\eps$ for general $W$.  However, if one
improves them, then we automatically get an improved upper bound on the
running time of \emph{the same} algorithm.

Henceforth, let $r, s$ be sufficiently small.  The
average probability over all sets $W$ with $t$ ones is $\Exp[_{W: |W| = t}]
{\eps(W,r,s)} = \Om{|W| \epsrs}$ (Lemma~\ref{lem:exp-prob}).  We are able to
prove the same bound for all $|W| \le \sqrt n$ (Lemma~\ref{lem:small-set}), when $W$ is
an independent set, that is it does not contain two ones in the same row or column
(Lemma~\ref{lem:indep-set}), or when the ones in $W$ form a rectangle
(again Lemma~\ref{lem:small-set}).  However, the latter rectangle bound
only holds for a limited class of $r, s$, which does not include the
balanced case $r = s = k$ in the range used by our algorithm.  As a
consequence, if the ones in $W$ form a full row or a column, our algorithm is
slower than what would follow from this formula.  We, however, show that in
this case our algorithm is optimal (Theorem~\ref{th:lowerb-verify}); this is
the only known tight lower bound for our algorithm.  Most of the proofs are
postponed to Appendix~\ref{app:fraction-marked}.

\begin{Lemma}
\label{lem:exp-prob}
Let $r s \le \frac {n^2} t$.  Then $\Exp[_{W: |W| = t}] {\eps(W,r,s)} =
\Om{|W| \epsrs}$.
\end{Lemma}

\begin{Lemma}
\label{lem:small-set}
Let $w \rem{= \# \{ i \given \exists j: W_{i,j} \ne 0 \}}$ be the number of
nonzero rows of $W$, and let $w'$ be maximal
the number of nonzero entries in a row.  Then for every $r \le \frac
n w$ and $s \le \frac n {w'}$, $\eps(W,r,s) = \Om{|W| \epsrs}$.
\end{Lemma}

\begin{Lemma}
\label{lem:indep-set}
Let $W$ have at most one entry in every row and column.  Then for every $r, s$
satisfying $r s \le {n^{4/3} / |W|^{2/3}}$, $\eps(W,r,s) = \Om{|W| \epsrs}$.
\end{Lemma}

The main Lemma~\ref{lem:eps} is a direct corollary of
Lemmas~\ref{lem:small-set} and~\ref{lem:indep-set}.

\medskip
\begin{Proof}[Proof of Lemma~\ref{lem:eps}.]%
Lemma~\ref{lem:small-set} implies that $\eps(W,k) = \Om{\frac {k^2} {n^2} \min
(|W|, \sqrt n)}$:  First, assume that $|W| \le \sqrt n$
and verify the restrictions on $r=s=k$.  For every $t \le \sqrt n$
it holds that $n^{2/3} / t^{1/3} \le n / t$.  Hence if $|W| \le \sqrt n$, then
for every $k \le n^{2/3} / |W|^{1/3}$ it holds that $k \le n / |W|$ and,
since $w, w' \le |W|$, also $k \le \frac n w$ and $k \le \frac n
{w'}$.  Hence the lower bound $\eps(W,k) = \Om{\frac {k^2} {n^2} |W|}$
given by Lemma~\ref{lem:small-set} holds for every $k$ in the range required
by Lemma~\ref{lem:eps}.  Now, if $|W| > \sqrt n$, the bound follows from
the monotonicity of $\eps(W,k)$ in $W$.

\rem{
This implies a stronger lower bound for arbitrary $W$ of size larger than
$n^{3/2}$, because every $W$ contains an independent set of size at least $|W|
/ n$.
}

Lemma~\ref{lem:indep-set} says that $\eps(W',k) = \Om{\frac {k^2} {n^2} |W'|}$
for every independent $W'$ and $k$ in the range required by
Lemma~\ref{lem:eps}.  The bound on $W$ follows from the monotonicity of
$\eps(W,k)$ in $W$.
If we put these two bounds together, we obtain that $\eps(W,k) = \Om{\frac
{k^2} {n^2} q(W)}$, as desired.
\end{Proof}

The bound cannot be strengthened to $\eps(W,k) = \Om{\frac {k^2} {n^2} |W|}$
for general $W$ and full range of $k$.  We show that \emph{no} quantum algorithm can be
fast if the $n$ ones in $W$ form a full row.  A straightforward calculation shows
that $q(W)$ for this $W$ can be at most $\sqrt n$ if we want the bound on
$\eps$ to hold for all $k \le \O{n^{2/3} / q(W)^{1/3}}$.
\rem{
Consider a $W$ of size $n$ that contains all its entries in some fixed
row, and let $|R| = k$.  Clearly, $\nonempty$ iff $W \subseteq R$.
Hence $\eps(W,k) = \Pr \nonempty = \Pr{W \subseteq R} = \frac k n$, the
number of iterations of the quantum walk is $\ell = 1 / \sqrt{ \delta_G \,
\eps(W,k) } = \sqrt n$ since $\delta_G = \Th{\frac 1 k}$, and the running
time of {\em Product Verification} is $\Th{n^{3/2}}$.
}


\begin{Theorem}
\label{th:lowerb-verify}
Any bounded-error quantum algorithm distinguishing a correct matrix product
and a matrix product with one wrong row has query complexity
$\Om{n^{3/2}}$.
\end{Theorem}

\begin{Proof}
We reduce Or of $n$ parities of length $n+1$ to product verification.  Let
\[
z = (x_{1,1} \oplus \dots \oplus x_{1,n} \oplus y_1) \vee \dots \vee
  (x_{n,1} \oplus \dots \oplus x_{n,n} \oplus y_n).
\]
Using the quantum adversary lower bound method~\cite{ambainis:lowerb}, it
follows that computing $z$ requires $\Om{n^{3/2}}$
quantum queries, and the lower bounds holds even if we promise that at most
one parity is equal to 1.  Since $z=1$ iff $\exists i: y_i \ne
\bigoplus_{\ell=1}^n x_{i,\ell}$, we can reduce this problem to the
verification of the matrix product $A B = C$ over $\GF 2$, where $A_{i,j} =
x_{i,j}$, $B_{i,j} = 1$, and $C_{i,j} = y_i$.  The promise is transformed into
that at most one row is wrong.
\end{Proof}

\section{Concluding remarks}

\subsection{Algorithm for computation of matrix products}

Let $m \ge n^{2/3}$.
One can modify the algorithm to verify the product $A_{n \times m}
B_{m \times n} = C_{n \times n}$ in time proportional to $n^{2/3} m$.
The quantum walk stays the same and only the inner scalar
products are of length $m$ instead of $n$.

Using the rectangular product verification algorithm and binary search,
one can construct a quantum algorithm that outputs
the position of a wrong entry.  By iterating this and correcting the wrong
entries, one can compute the matrix product $A B = C$ whenever a good
approximation to $C$ is known.  One can always start by guessing $C = 0$, hence
the following bound holds:

\begin{Theorem}
\label{th:compute}
Let $m \ge n^{2/3}$.
The matrix product $A_{n \times m} B_{m \times n} = C_{n \times n}$ can be
computed with polynomially small error probability in expected time
\begin{equation}
\label{eq:matrix-mult}
T_M \le \O 1 \cdot \left\{      \begin{array}{l l l}
m \log n \cdot n^{2/3} w^{2/3};       & \mbox{for} & 1 \le w \le \sqrt n, \\
m \log n \cdot \sqrt n w;             & & \sqrt n \le w \le n, \\
m \log n \cdot n \sqrt w;             & & n \le w \le n^2,
\end{array} \right.
\end{equation}
where $w = |W|$ is the number of nonzero entries of $C$.
\end{Theorem}

The algorithm and its analysis are presented in Appendix~\ref{app:compute}.
Let us neglect the logarithmic term.  It follows that matrix products with
$|W| = o(\sqrt n)$ non-zero entries can be computed in sub-quadratic time $o(n
m)$.  We can also compare our algorithm to the best classical
algorithms, however this comparison cannot be fair, since our algorithm
depends on $|W|$, whereas all known classical algorithms depend on the sparseness
of the input matrices.  The fastest known algorithm for dense square
matrices~\cite{cw:matrix-mult} works in time $\O{n^{2.376}}$.  Our algorithm
can beat it when the number of nonzero elements of the result is $|W| =
o(n^{0.876})$.  The fastest known algorithm for dense rectangular
matrices~\cite{coppersmith:matrix} works in time $\O{n^{1.844 + o(1)}
m^{0.533} + n^{2 + o(1)}}$.  The fastest known algorithm for sparse square
matrices~\cite{yz:matrix} works in time $\O{n^{1.2} z^{0.7} + n^{2 + o(1)}}$,
where $A$ and $B$ have at most $z$ non-zero elements.

\subsection{Boolean matrices}

The algorithm \VO{}
relies on the fact that arithmetical operations are over some integral
domain.
If the matrices are over the Boolean algebra $\{ \vee, \& \}$, then the
multiplication by random vectors from both sides does not work.
However, Boolean matrix products can be verified even faster by the following
algorithm:

\begin{Theorem}
\label{th:verify-bool}
There exists a quantum Boolean-matrix product verification algorithm running in time
$\O{n \sqrt m}$ and space $\O{\log n + \log m}$.
\end{Theorem}

\begin{Proof}
The condition that three given matrices form a valid product can be written as
an And-Or tree: And of $n^2$ equalities, each being an Or of $m$ products.
There is a bounded-error quantum algorithm~\cite{hmw:berror-search}
running in time $\O{\sqrt{n^2 m}} = \O{n \sqrt m}$ and space $\O{\log (n^2 m)}$.
\end{Proof}

By standard techniques~\cite{bbht:bounds}, one can speed up the verification
to time $\O{n \sqrt{m / t}}$, if the number of wrong entries $t$ is known
beforehand.  If $t$ is unknown, then the verification can be done in
\emph{expected time} $\O{n \sqrt{m / t}}$ and the worst-case time stays $\O{n
\sqrt m}$.  The Boolean matrix product with $t$ nonzero entries can be thus
computed in expected time $\O{n \sqrt{t m}}$.

\subsection{Open problems}

It would be interesting to strengthen the lower bound on the fraction
$\eps(W,k)$ of
marked pairs and thus also the upper bound on product verification.  As we
have shown, this cannot be done in full generality, but perhaps one can show a
stronger lower bound using some density argument.

The time complexity of our algorithm goes up if the
space complexity is bounded.  Can one prove a time-space tradeoff for the
verification problem similar to the tradeoff for computation of matrix
products~\cite{ksw:dpt}?  Note that we currently can't show time-space
tradeoffs for {\em any} decision problem.

Can one prove a better lower bound on verification of matrix products than
$\Om{n^{3/2}}$?  This lower bound is tight when there are $\sqrt n$ wrong
entries.  Is the true bound higher with only one wrong entry?
Due to the small certificate complexity of this problem, one cannot prove such
a bound using any of the adversary methods~\cite{ss:adversary}, but it might
be provable by the polynomial method~\cite{bbcmw:polynomialsj}.

\section*{Acknowledgments}
We thank Ronald de Wolf and Troy Lee for useful discussions.
We thank anonymous referees for their valuable comments.

\bibliographystyle{alpha}
\bibliography{../quantum}
\appendix

\section{Proofs for the quantum walk}
\label{app:qwalk}

\begin{Proof}[Proof of Theorem~\ref{th:q-walk}.]
This is a corollary of Lemma~7 from~\cite{szegedy:q-walk}.  To
express the lower bound on $m$ in terms of $\delta_G, \eps$, we use several
other statements from that paper: Let $P = {\cal L}(G)$ be the Laplacian of
$G$ and let $P_M$ be obtained from $P$ by leaving out all rows and columns
indexed by some $x \in M$.  By~\cite[Lemma~10]{szegedy:q-walk}, $\lambda(P_M) \le 1 - \delta_G
\eps /2$.  The lower bound on $m$ can be restated as $\Om{\sqrt{1
\over 1 - \lambda(P_M)}} = \Om{1/\sqrt{\delta_G \eps}}$ like in
\cite[Corollary~2]{szegedy:q-walk}.
\end{Proof}

\begin{Proof}[Proof of Lemma~\ref{lem:spectral-gap}.]
It is not difficult to show that the spectral gap of the Johnson graph
$J(n,k)$ is ${n \over (n-k) k}$, which is $\Th{1/k}$ for $1 \le k
\le \frac n 2$.  Furthermore, it is simple to prove that $\delta_{G_1 \times
G_2} = \min( \delta_{G_1}, \delta_{G_2} )$.
We conclude that $\delta_G = \Th{1/k}$.
\end{Proof}

\rem{
The following standard lemma finishes the analysis of \VO{}.  We
plug the uniform superposition as $\ket \varphi$ and the superposition
computed by the quantum walk as $\ket \psi$.
}

\begin{Proof}[Proof of Lemma~\ref{lem:test}.]
$\ket Y = \frac 1 2 (\ket{0, \varphi} + \ket{1, \varphi}) + \frac 1 2 (\ket{0,
\psi} - \ket{1, \psi}) = \ket 0 {\ket \varphi + \ket \psi \over 2} + \ket 1
{\ket \varphi - \ket \psi \over 2}$, hence $\Pr{Y = 1} = \norm{ \ket \varphi -
\ket \psi \over 2}^2 = \frac 1 4 (\bra \varphi - \bra \psi)(\ket \varphi -
\ket \psi) = \frac 1 4 (\braket \varphi \varphi + \braket \psi \psi - 2
\braket \varphi \psi) = \frac 1 2 (1 - \braket \varphi \psi)$.
\end{Proof}

\section{Proofs for the fraction of revealing pairs}
\label{app:fraction-revealing}

\begin{Lemma}
\label{lem:int-dom}
Let $G$ be an integral domain with $g$ elements, let $(R,S)$ be marked, and
let $\Vp, \Vq$ be picked uniformly at random from $G^n$.  The probability that
$(R,S)$ is revealing is $\ge (1 - \frac 1 g)^2 \ge \frac 1 4$.
\end{Lemma}

\begin{Proof}
Let $D = A B - C$, that is the wrong entries are exactly the nonzero entries
of $D$.  Assume that $(R, S)$ is marked and pick any $D_{i_0,j_0}
\ne 0$.  Now, $(R,S)$ is not revealing iff
\[
0
  = \sum_{i \in R, j \in S} \Vp_i \Vq_j D_{i,j}
  = \sum_{i \in R, j \in S \atop D_{i,j} \ne 0} \Vp_i \Vq_j D_{i,j}
  = \Vp_{i_0} (\Vq_{j_0} D_{i_0,j_0} + c_1) + c_2 \Vq_{j_0} + c_3,
\]
where $c_1, c_2, c_3$ are some constants depending on $D$ and other
coordinates of $\Vp, \Vq$.  Fix these constants and pick $\Vp_{i_0}, \Vq_{j_0}$
at random from $G$.  Since $G$ is an integral domain with $g$ elements,
$p = \Pr{\Vq_{j_0} D_{i_0,j_0} + c_1 = 0} \le \frac 1 g$.
For every $\Vq_{j_0}$ such that $\Vq_{j_0} D_{i_0,j_0} +
c_1 \ne 0$, the equality is equivalent to $c_4 \Vp_{i_0} + c_5 = 0$ for another
constants $c_4 \ne 0$ and $c_5$, which is again satisfied by at most $1$ value of
$\Vp_{i_0} \in G$.  Hence the probability of having equality by chance is
at most
\[
p \cdot 1
  + (1 - p) \cdot \frac 1 g
  = \frac 1 g + p \cdot \frac {g-1} g
  \le \frac 1 g + \frac {g-1} {g^2}
  = \frac {2 g -1} {g^2}.
\]
The probability that $(R,S)$ is revealing is thus at least $1 - \frac
{2g-1} {g^2} = (1 - \frac 1 g)^2 \ge \frac 1 4$ and the equality holds when
$g = 2$.
\end{Proof}

\begin{Lemma}
\label{lem:expect}
Let $0 \le X \le 1$ and $\Exp X \ge \alpha$.
Then $\Pr{ X \ge \beta } > \alpha - \beta$.
\end{Lemma}

\begin{Proof}
Decompose the expected value of $X$ conditioned on $X \ge \beta$:
\[
\Exp X = \Exp{X | X < \beta} \cdot (1 - \Pr{X \ge \beta})
  + \Exp{X | X \ge \beta} \cdot \Pr{X \ge \beta}.
\]
Rearrange, plug in $0 \le X \le 1$, and obtain
\(
\Pr{X \ge \beta}
  = {\Exp X - \Exp {X | X < \beta} \over
    \Exp{X | X \ge \beta} - \Exp{X | X < \beta}}
  > { \alpha - \beta \over 1 - 0}
  = \alpha - \beta.
\)
\end{Proof}

\begin{Proof}[Proof of Lemma~\ref{lem:good-vec}.]
Consider Boolean random variables $V_{R,S,\Vp,\Vq} = 1$ iff $(R,S,\Vp,\Vq)$ is
revealing.  Let $v_{\Vp,\Vq}$ be the \emph{fraction of marked sets that are also
revealing} when $\Vp,\Vq$ multiply the equation, formally $v_{\Vp,\Vq} =
\Exp[_{{\rm marked}\, (R,S)}]{V_{R,S,\Vp,\Vq}}$.
By Lemma~\ref{lem:int-dom}, for every marked $(R, S)$,
$\Exp[_{\Vp,\Vq}]{V_{R,S,\Vp,\Vq}} \ge \frac 1 4$.
It follows that
$\Exp[_{{\rm marked}\, (R,S)}] {\Exp[_{\Vp,\Vq}]{V_{R,S,\Vp,\Vq}}} \ge \frac 1 4$
and hence
$\Exp[_{\Vp,\Vq}] {\Exp[_{{\rm marked}\, (R,S)}] {V_{R,S,\Vp,\Vq}} }
= \Exp[_{\Vp,\Vq}]{ v_{\Vp,\Vq} } \ge \frac 1 4$.
By Lemma~\ref{lem:expect}, when $\Vp, \Vq$ is picked uniformly at random,
$\Pr{ v_{\Vp,\Vq} \ge \frac 1 8} > \frac 1 8$.
Hence in this lucky case, $\zeta_{\Vp,\Vq}(W,k) \ge \frac 1 8 \eps(W,k)$.
\end{Proof}

\section{Proofs for the fraction of marked pairs}
\label{app:fraction-marked}

\begin{Proof}[Proof of Lemma~\ref{lem:exp-prob}.]
Consider Boolean random variables $V_{R,S,W} = 1$ iff $\nonempty$.  Then for
every $|R| = r$ and $|S| = s$, it holds that $\Exp[_{W: |W| = t}]{ V_{R,S,W} }
= \Pr[_W] \nonempty$ and
\begin{eqnarray*}
\Pr[_{W: |W| = t}] \nonempty
  &=& 1 - {n^2 - r s \over n^2} \cdot {n^2 - r s - 1\over n^2 - 1}
    \cdots {n^2 - r s - t + 1 \over n^2 - t + 1} \\
&\ge& 1 - \left({ n^2 - r s \over n^2 }\right)^t
  = 1 - \left(1 - {r s \over n^2 }\right)^t
  \ge 1 - e^{-{r s \over n^2 } t}
  = \Omega\left( t \epsrs \right),
\end{eqnarray*}
because $1 - x \le e^{-x}$ and, on any fixed interval $x \in [0,A]$, also
$e^{-x} \le 1 - {1 - e^{-A} \over A} x$.  The claim is now proved using
standard arguments.  Since $\forall R, S: \Exp[_W] {V_{R,S,W}} \ge t
\epsrs$, also $\Exp[_{R,S}] {\Exp[_W] {V_{R,S,W}}} \ge t \epsrs$.
Exchange the order of summation and obtain $\Exp[_W] {\Exp[_{R,S}]
{V_{R,S,W}}} = \Exp[_W] {\eps(W,{r,s})} \ge t \epsrs$.
\end{Proof}

\begin{Proof}[Proof of Lemma~\ref{lem:small-set}.]
Let $Z$ denote the random event ``$\nonempty$''.  For $j = 0, 1, \dots, w$,
let $Z_j$ denote the random event ``$W \cap R \times [n]$ has exactly $j$ nonempty
rows''.  Since $\{ Z_j \}$ are disjoint and $\sum_{j=0}^w \Pr{Z_j} = 1$, we
can decompose the probability
\[
\Pr Z = \sum_{j=0}^w \Pr{Z \given Z_j} \cdot \Pr{Z_j}
  \ge \sum_{j=1}^w \Pr{Z_j} \cdot \Pr{Z \given Z_1}
  = (1 - \Pr{Z_0}) \cdot \Pr{Z \given Z_1},
\]
because $\Pr{Z \given Z_j} \ge \Pr{Z \given Z_1}$ for $j \ge 1$.
Now, $\Pr{Z_0} = \Pr{W \cap R \times [n] = \emptyset} = \frac {n-w} n \cdot \frac {n-w-1}
{n-1} \cdots \frac {n-w-r+1} {n-r+1} \le (\frac {n-w} n)^r = (1 - \frac w n)
^r \le e^{- \frac {r w} n}$, because $1 - x \le e^{-x}$.  Recall that for every
$x \in [0,A]$, $e^{-x} \le 1 - {1 - e^{-A} \over A} x$.  If $r \le \frac n w$,
then $\frac {r w} n \le 1$ and hence $e^{-\frac {r w} n} \le 1 - {1 - e^{-1}
\over 1} \frac {r w} n = 1 - \alpha \frac {r w} n$ for $\alpha = 1 - e^{-1}$.  We conclude that
$1 - \Pr{Z_0} \ge \alpha \frac {r w} n$.

To lower-bound the other term, we decompose $Z_1$.  For $i = 1, 2, \dots, n$,
let $Y_i$ denote the random event ``$W \cap R \times [n]$ has the $i$-th row nonempty and
all other rows are empty''.  Let $w_i$ be the number of entries in the $i$-th
row of $W$ and let $w' = \max_i w_i$.  Since $\{ Y_i \}$ are disjoint and $Y_1 \cup
\dots \cup Y_n = Z_1$,
\[
\Pr{Z \given Z_1}
  = \sum_{i: w_i \ne 0} \Pr{Z \given Y_i} \cdot \Pr{Y_i \given Z_1}
  = \frac 1 w \sum_{i: w_i \ne 0} \Pr{Z \given Y_i}.
\]
$\Pr{Z \given Y_i}$ is easy to evaluate, since the $i$-th row of $W$ contains
exactly $w_i$ entries and $S$ is picked uniformly at random.  Let $W_i =
W \cap R \times [n]$ be the $i$-th row of $W$.  By the same arguments as above, $\Pr{Z
\given Y_i} = \Pr{W_i \cap [n] \times S \ne \emptyset \given |W_i| = w_i} = 1 - \Pr{W_i
\cap [n] \times S = \emptyset \given |W_i| = w_i} \ge 1 - e^{- \frac {s w_i} n}$.
Analogously, if $s \le \frac n {w_i}$, then $e^{-{s w_i \over n}} \le 1 -
\alpha \frac {s w_i} n$ and $\Pr{Z \given Y_i} \ge \alpha \frac {s w_i} n$.
Plug both bounds together and obtain $\Pr Z \ge \alpha \frac {r w} n
\frac 1 w \sum_{i: w_i \ne 0} \alpha \frac {s w_i} n = \alpha^2 \frac
{r s} {n^2} \sum_{i: w_i \ne 0} w_i = \Om{|W| \epsrs}$, as desired.
\end{Proof}

\begin{Proof}[Proof of Lemma~\ref{lem:indep-set}.]
If $t \le \sqrt n$, then the result follows from Lemma~\ref{lem:small-set}.
Let us assume that $t > \sqrt n$.  Again, let $Z$ denote the random event
``$\nonempty$'' and, for $j = 0, 1, \dots, r$, let $Z_j$ denote the random
event ``$W \cap R \times [n]$ has exactly $j$ nonempty rows''.  Then
\[
1 - \Pr{Z \given Z_j}
  = \Pr{W \cap R \times S = \emptyset \given Z_j}
  = \frac {n-s} n \cdot \frac {n-s-1} {n-1} \cdots \frac {n-s-j+1} {n-j+1}
  \le \left({ n-s \over n }\right)^j
  \le e^{- {s j \over n}}.
\]
Since $j \le r$ and $t > \sqrt n$, we get that $s j \le r s \le n^{4/3} /
t^{2/3} \le n^{4/3} / (\sqrt n)^{2/3} = n$.  Hence $\frac {s j} n \le 1$ and
by upper-bounding the exponential we get that $\Pr{Z \given Z_j} \ge 1 - e^{-
{s j \over n}} \ge 1 - (1 - \alpha {s j \over n}) = \alpha {s j \over n}$ for
$\alpha = 1 - e^{-1}$.
Now, $\{ Z_j \}$ are disjoint and $\sum_{j=0}^r \Pr{Z_j} = 1$, hence we can
decompose the probability
\[
\Pr Z = \sum_{j=0}^r \Pr{Z \given Z_j} \cdot \Pr{Z_j}
  \ge \sum_{j=0}^r \alpha \frac {s j} n \Pr{Z_j}
  = \alpha \frac s n \sum_{j=0}^r j \cdot \Pr{Z_j}
  = \alpha \frac s n \cdot \Exp{Y},
\]
where $Y$ is the number of nonempty rows.
There are $r$ rows among $n$ in $R$ and we pick $t$ entries without
returning uniformly at random.  An application of $\Exp{Y} = \frac {r
t} n$ completes the proof.
\end{Proof}

\section{Computation of matrix products}
\label{app:compute}

In this section, we show how to use (the rectangular version of) \emph{Product Verification} to obtain the
actual position of a wrong entry.  Furthermore, we present an algorithm for
computation of matrix products.  The algorithms are described in
Figure~\ref{fig:compute}.

\begin{figure}[tbp]
\begin{center}
\framebox{
\begin{minipage}{0.9\hsize}
\noindent
{\em Matrix Multiplication}
(input size $n, m$, matrices $A_{n \times m}, B_{m \times n}$)
returns $C_{n \times n} = A B$:
\begin{shortenum}
\item
Initialize $C = 0$.
\item
\label{step:find-wrong}
Run {\em Find Wrong Entry} $(n, m, A, B, C)$. \\
If it returns ``equal'', return $C$.
\item
Otherwise let $(r,c)$ be the wrong position.  Recompute $C_{r,c}$. \\
Find and recompute all wrong entries in the $r$-th row using the {\em Grover Search}.  \\
Find and recompute all wrong entries in the $c$-th column using the {\em Grover Search}.
\item
Go to step~\ref{step:find-wrong}.
\end{shortenum}
\end{minipage}
}
\vskip 10pt
\framebox{
\begin{minipage}{0.9\hsize}
\noindent
{\em Find Wrong Entry}
(input size $n, m$, matrices $A_{n \times m}, B_{m \times n}, C_{n \times n}$)
\\
returns a position $(r,c)$ if $C_{r,c} \ne \sum_i A_{r,i} B_{i,c}$ or ``equal'' if $A B = C$:
\begin{enumerate}
\item
If $n=1$, verify the scalar product and exit.
\item
Let $A_1, A_2$ denote the top and bottom half of $A$, \\
let $B_1, B_2$ denote the left and right half of $B$, and \\
let $C_{1,1}, C_{1,2}, C_{2,1}, C_{2,2}$ denote the four quadrants of $C$.
\item
Repeat at most $\O{\log n}$ times the following step:
\begin{shortitem}
\item
Run in parallel {\em Product Verification} $(\frac n 2, m, A_i, B_j,
C_{i,j})$ for $i, j \in \{ 1, 2 \}$. \\
If some of them returns ``not equal'', stop the other threads of computation
and cancel the loop.
\end{shortitem}
\item
If the product verification was always successful, return ``equal''.
\item
Let $C_{i,j} \ne A_i B_j$ be the found wrong sub-matrix. \\
Let $(r',c') = $ {\em Find Wrong Entry} $(\frac n 2, m, A_i, B_j, C_{i,j})$.
\item
If $i=1$, set $r = r'$, otherwise set $r = r' + \frac n 2$. \\
If $j=1$, set $c = c'$, otherwise set $c = c' + \frac n 2$. \\
Return $(r,c)$.
\end{enumerate}
\end{minipage}
}
\end{center}
\caption{Quantum algorithm for computation of matrix products}
\label{fig:compute}
\end{figure}

\begin{Theorem}
\emph{Find Wrong Entry} has one-sided polynomially small error,
worst-case running time $\O{n^{2/3} m \log n}$, and expected running time
$\O{n^{2/3} m \log n / q(W)^{1/3}}$ for the set of wrong entries $W$.
\end{Theorem}

\begin{Proof}
Assume that $A B \ne C$.  Let $W^\ell$ be the set of wrong entries in the
$\ell$-th recursion level of the binary search.  From the definition of $q(W)$, if
$q(W^\ell) = q$, then $q(W^\ell_{i,j}) \ge \frac q 4$ for at least one
quadrant $i, j \in \{ 1, 2 \}$.  {\em Find Wrong Entry} descends into the
first quadrant it finds a solution in, hence it chooses $W^{\ell+1} =
W^\ell_{i,j}$ with high probability and then $(\frac n
2)^2 / q(W^{\ell+1}) \le n^2 / q(W^\ell)$.  There are $\log n$ levels of the
recursion.  Hence its expected running time is at most
\[
\sum_{\ell=1}^{\log n} 4 \sqrt[3]{(n / {2^\ell})^2 \over q(W^\ell)} m
  \le \sum_{\ell=1}^{\log n} 4 \sqrt[3]{n^2 \over q(W)} m
  = {n^{2/3} m \over q(W)^{1/3}} \sum_{\ell=1}^{\log n} 4
  = \OO{{n^{2/3} m \over q(W)^{1/3}} \log n},
\]
as claimed.  By Theorem~\ref{th:exp-time}, the probability that a wrong matrix
product is not recognized in one iteration is at most $\frac 1 3$.  The
probability that it is not recognized in $\O{\log n}$ iterations is $1 /
\poly(n)$.  If $A B = C$, then the first iteration of binary search is
repeated $\O{\log n}$ times and the worst-case running time is $\O{n^{2/3} m
\log n}$.
\end{Proof}

\begin{Remark}
It might be that the position of the wrong entry can be obtained from just one run of {\em
Product Verification} in the same way as in the quantum walk algorithm for
element distinctness~\cite{ambainis:eldist} -- by measuring the subsets $R,S$
instead of the quantum coin register $\ket z$.  However, this is only known
to follow from Theorem~\ref{th:q-walk} for exactly one wrong entry, that is 
$|W|=1$~\cite[Section 10]{szegedy:q-walk}.  The log-factor in the total
running time is necessary for polynomially small error.
\end{Remark}

Now we can prove the upper bound on matrix multiplication.
\medskip

\begin{Proof}[Proof of Theorem~\ref{th:compute}.]%
Finding all $r_\ell$ wrong entries in the $\ell$-th row is done by the Grover
search with unknown number of solutions~\cite{bbht:bounds}, and it takes time
\(
\sum_{i=1}^{r_\ell} \sqrt{n \over i} m
  = \O{\sqrt{n r_\ell} m},
\)
where the scalar products of length $m$ are computed on-line.  We ensure that
there are no wrong entries left with probability polynomially close to one in
additional time $\O{\sqrt n m \log n}$.  Let us condition the rest of the
analysis by that the Grover searches indeed find all ones.

Let $W'$ be the largest independent subset of $W$.  Clearly, {\em Matrix
Multiplication} finishes in at most $|W'|$ iterations, otherwise there would
exist an independent set larger than $|W'|$.  The total running time is the
sum of the time spent in {\em Find Wrong Entry}
\[
T_F \le \sum_{\ell=1}^{|W'|} {n^{2/3} m \log n \over |W'|^{1/3}}
  = \O{(n |W'|)^{2/3} m \log n},
\]
and the time spent in the Grover searches.  By applying a Cauchy-Schwarz
inequality several times,
\begin{eqnarray*}
T_G &\le& \sum_{\ell=1}^{|W'|} \left(
    \sqrt{n r_\ell} m + \sqrt{n c_\ell} m \right) \log n
  = m \sqrt n \log n \left( \sum_{\ell=1}^{|W'|} 1 \cdot \sqrt{r_\ell}
    + \sum_{\ell=1}^{|W'|} 1 \cdot \sqrt{c_\ell} \right) \\
&\le& m \sqrt n \log n \sqrt{ \sum_{\ell=1}^{|W'|} 1 }
    \left( \sqrt{\sum_{\ell=1}^{|W'|} r_\ell}
      + \sqrt{\sum_{\ell=1}^{|W'|} c_\ell} \right)
    = \O{ m \sqrt n \log n \sqrt{|W'|} \sqrt{|W|}}.
\end{eqnarray*}
The algorithm is
bounded-error, because both {\em Find Wrong Entry} and the iterated Grover
searches have polynomially small error.  Put the bounds together and obtain:
\[
T_M = T_F + T_G \le
  m \log n \sqrt n \sqrt{|W'|} \cdot ( n^{1/6} |W'|^{1/6} + \sqrt{|W|}).
\]
Evaluate separately the three cases $|W| \in [1,\sqrt n]$, $|W| \in [\sqrt n, n]$,
and $|W| \in [n, n^2]$, use that $|W'| \le |W|$ and $|W'| \le n$, and obtain
inequality~(\ref{eq:matrix-mult}), which we had to prove.
\end{Proof}


\end{document}